# Optimizing Location Allocation in Urban Management: A Brief Review

Aref Ayati, Mohammad Mahdi Hashemi, Mohsen Saffar, Hamid Reza Naji
*dept. Computer Engineering and Information Technology*
*Graduate University of Advanced Technology*
Kerman, Iran
a.ayati@kgut.ac.ir, mohammad.hashemi@student.kgut.ac.ir, mohsen.saffar@email.kntu.ac.ir, Naji@kgut.ac.ir

*Abstract*— Regarding the concepts of urban management, digital transformation, and smart cities, various issues are presented. Currently, we like to attend to location allocation problems that can be a new part of digital transformation in urban management (such as locating and placing facilities, locating and arranging centers such as aid and rescue centers, or even postal hubs, telecommunications, electronic equipment, and data centers, and routing in transportation optimization). These issues, which are seemingly simple but in practice complex, are important in urban environments, and the issue of accurate location allocation based on existing criteria directly impacts cost management, profit, efficiency, and citizen satisfaction. In recent years, researchers have used or presented various models and methods for location allocation problems, some of which will be mentioned in this article. Given the nature of these problems, which are optimization problems, this article will also examine existing research from an optimization perspective in summary. Finally, a brief conclusion will be made of the existing methods and their weaknesses, and suggestions will be made for continuing the path and improving scientific and practical research in this field.

## I. Introduction

As a notion of digital transformation, digitalization is frequently linked to digitization activities including data exchange, business process improvement, and internal workflow optimization [1]. Digital transformation refers to the integration and exploitation of digital technologies to increase productivity and social well-being [2]. Manufacturing processes are undergoing a fundamental transformation due to automation and information technology. More competitive and efficient manufacturing is made possible by the digitization and optimization of processes [3]. One aspect of optimization in smart cities is optimization in location allocation issues that can go a city smarten and more digitalized than before.

Location modeling is a method for optimizing the location of facilities or equipment and the allocation of consumers or demands to facilities [4]. This modeling system is widely used for facility location planning in both the public and private sectors [5]. Facilities that offer any kind of service must be located in a way that reduces social costs and enhances benefits to stakeholders by following certain guidelines. The location problem is an NP-HARD problem that becomes exponentially more complex as the number of points required for location increases and the number of service recipients increases, which means that exact and mathematical techniques cannot be applied in all cases, so depending on their structure and subject, heuristic methods must also be used. This problem was raised by [6] and researchers stated that it is better to solve most location and location problems based on new optimization methods, which are often heuristic or meta-heuristic. NP problems are complex problems in the field of decision-making that can be solved in polynomial time using a non-deterministic Turing machine. The most complex problems in this field are called NP-Hard [7]. On the other hand, the problem of facility or equipment location is a part of operations research that is of great importance for both practical and combinatorial optimization perspectives, as well as classical location problems related to determining the location of a specific facility to optimize the allocation of facilities to customers [8]. In recent years, facility location problems, specializing in location allocation , have been very widespread, and just as examples, we can mention cases such as locating the location of urban facilities from metro stations, taxi, bus, and bicycle stations to garbage bins, locating the location of relief and crisis centers, locating the location of telecommunication facilities such as Base Transceivers Stations (BTS) antennas, locating the location of hubs in the case of shipment transfer in the supply chain, locating military centers and so on.

Researchers have previously tried to solve problems like this using mathematical methods as optimally as possible, but with the advent of heuristic and meta-heuristic algorithms in the field of soft computing, we are experiencing a new era in the field of modern optimization problems. In the field of optimization problems in the field of location allocation , some problems need to be investigated critically, accurately, and as quickly as possible, while for some other problems, a slower speed of conclusion is required. On the other hand, the type of calculation of each problem and the algorithm used can also cause a change in the conclusion. For example, given that in 2019, we witnessed more severe disasters than before, causing more deaths than any year in the previous decade, and based on the statistical data of natural disasters reported by the Center for Research on Epidemiology of Natural Disasters, floods were the most common disaster, causing 5,110 deaths and economic losses of $36.8 billion [9]. This forces researchers to prioritize issues that involve human lives, such as allocating space for safe havens during a disaster over allocating space for bicycle stations. However, all of these issues need to be addressed in some way.

In the field of digital transformation and industry 4, especially in the area of Lean 4 management, the requirement to minimize the three factors of waste or Muda, inconsistency or Mura, and overburden or Muri is of great interest [10]. By using optimization in the field of location allocation or routing in the field of urban management, these three factors can be greatly reduced. In the remainder of this article, we will review the methods presented in various articles in recent years, referring to the topics and the results they have obtained on these fields for location allocation, trying to simple analysis, and finally, give some suggestions for future work will be presented.

## II. BACKGROUND

### A. Optimization

Resources are always limited, and optimizing the available resources is very important. Optimization which is one of the key factors of Industry 4.0 can be used in various fields, from engineering design to economics or planning a trip or computer network's routing [11, 12].

In a proper optimization, using a process that is the result of mathematical models and computer algorithms, one can follow an objective function to discover the optimal response in a problem space. Many problems are generally continuous, discrete, bounded, or unbounded in nature [13, 14]. However, some problems are difficult to deal with using conventional mathematical methods such as conjugate gradient, fast steepest, and quasi-Newtonian methods [15]. Also, previously, simple optimization problems were classically discussed in the form of problems such as knapsack, n-queens, traveling salesman, and greedy problem-solving approaches, but currently, the use of heuristic and meta-heuristic methods such as nature-inspired algorithms are receiving more attention.

### B. Geospatial Information Systems (GIS):

Geographic information systems, or GIS, offer computer-based tools for storing, visualizing, analyzing, and interpreting geographic data. Geographic data (or spatial data) specifies the geographic location of each feature. This data includes anything that can be associated with a location on Earth, or more simply, anything that can be mapped. Roads, country borders, and addresses are all types of spatial data [16].

## III. SOME APPLICATIONS OF MODERN OPTIMIZATION METHODS FOR SOLVING SOME LOCATION ALLOCATION PROBLEMS

*1) Using a GIS-based spatial approach to determine the optimal locations of bikeshare stations: The case of Washington D.C*

The idea of bike sharing was explored in the 1960s and gained significant acceptance in the 1990s. It is now increasingly being pursued by users and policymakers, as the promotion of environmental models and sustainable urban mobility requires that efficient multimodal transport systems be made equally available to all users to foster outcomes such as cleaner living and the elimination of the need for cars [17, 18].

In the article [19], the authors initially chose Washington as a case study because it was one of the first cities in the US to consider planning the location of bike share stations and it was very well received so that this city eventually became the owner of one of the largest bike share systems in the country. Also, considering that one of the main elements in implementing a successful plan for locating bike stations was their integration with subway, bus, etc. stations, the researchers used the ArcGIS application to point out the current locations of the facilities and then created heat maps.

They then tried to establish a balance in the city's residential areas and areas where companies and work areas were located. In this article, modeling with two methods for optimization, namely MCMF and TMS (methods that are available in ArcGIS), has been used. These methods are not heuristic or meta-heuristic, and the first method is an approximate algorithm that tries to reduce the number of facilities by maximizing coverage, and the second algorithm is a greedy algorithm that tries to allocate in a way that preserves the market share of that area. It can be further stated that with the aforementioned method, the authors claim that their method can potentially be used to allocate optimal station locations in any city [19].

To optimize and streamline bike station location allocation, we can employ hybrid and current models with many techniques. Adding bike or bicycle stations to metro, transit, recreational, welfare, educational, and service facilities is a possibility. Bike station's location allocation models consider environmental variables, distance to critical centers, population grouping, government regulations, simplicity of use, and so on. Another method for solving problems with many, mutually exclusive goals is multi-objective optimization. These approaches are ideal for the optimum bicycle station location allocation issue, which concurrently analyzes objectives including greatest coverage, minimal distance, and other factors. These challenges can be solved using multi-objective optimization techniques like HGA, NSGAII, and others in future research and combination approaches. [20-23].

*2) Location allocation modeling for healthcare facility planning in Malaysia*

Nowadays, the issue of health facility location is of particular importance in all countries and regions. Now, their locations, their types, location conditions the level of their service, and the diversity of demand allocation to them, are of fundamental importance in order to maximize the benefits of the people and minimize social costs. In this context, the Maximal Covering Location Problem in One Place (MCLP) was presented in 1974 and has become one of the most common planning models. In this type of problem, the goal is to maximize the coverage. Especially in problems such as coverage of aid stations, fire stations, health stations, etc., where the tendency is to create maximum coverage with the least number of stations [24]. In recent years, however, MCLP has faced progress and can also solve larger problems with more than one objective and higher complexity [25, 26].

Later, however, CMCLP (Capacitated MCLP) methods were also proposed to include the issue of capacity, and in fact, CMCLP can be said to be an improvement over MCLP in which facilities have limited capacity and are unable to cover more than a certain limit of the population. In these problems, the goal is to maximize the population covered by a fixed number of facilities, with the condition that each population is assigned to its closest facility, and the capacity of the facilities is also respected [27].

In [28], the authors tried to use the CMCLP method and combine it with the genetic algorithm (GA) method for optimization. In this method, they first studied one of the places in Malaysia and studied its health facilities and then, using the GA, they created their unique method to solve the CMCLP problem because they included such problems in the class of CMCLP problems. In this method, they examined the percentage of population coverage by the current facilities within the maximum acceptable distance by the Malaysian government. Then, using CPLEX, which is a Python programming language library, they compared it on a medium-sized problem consisting of 179 network nodes and showed that their proposed algorithm produces better results. Then, they extended their method to a larger network consisting of 809 nodes, but CPLEX could not help them in this case. Overall, they showed that their proposed solution approach produces significant results in determining appropriate locations for facilities that maximize population coverage.

In location allocation optimization problems, binary approaches are common methods that require the transformation of the problem space into binary models. In the aforementioned research, a binary and extended version of the GA was used, and the approaches to transforming and modeling facilities in the form of binary solutions are innovations of the present research. The use of the GA in this research has shown positive results. To improve the quality of the solution and increase the validity of health facility location allocation problems, the use of other optimization algorithms such as ant colony optimization (ACO) is recommended due to the faster speed in reaching the optimal solution [29].

The use of other binary optimization methods and algorithms and comparison with the obtained results increases the quality of the results and the validity of scientific research in this field. Single-objective optimization algorithms are faster than multi-objective algorithms, but the use of multi-objective models ensures that solutions with different objectives and more valuable results are provided. In this regard, the multi-objective version of the GA can also be used to solve the location allocation problem [30].

*3) Optimization of facility location allocation model for base tranceiver station antenna establishment based on genetic algorithm considering network effectiveness criteria (case study north of Kermanshah)*

Given the widespread and intense need of mankind for communication, its uninterruptedness, and its maximum quality, the issue of locating base transceiver stations or BTSs has expanded greatly these days. These stations are places where antennas are installed to strengthen cellular computer network coverage in the area to increase the support capacity of cell phones. All cell phone operators also want to maximize their coverage for users while installing the least amount of facilities and equipment [31, 32].

In the process of predicting the location of telecommunication stations such as BTSs, however, very important things should be considered. For example, research that shows the effect of waves on the human body and also its effect on the mindset of people in a particular area, and at the same time the desire to increase coverage, reduce the number of facilities and the cost of setting up and providing services by service providers, and the like [33, 34], led the researchers in the present article [35] to desire to use an optimization method to investigate the best locations for BTS placement.

In this paper, researchers have used the Analytical Hierarchy Process (AHP), a method for decision-making. Using this method, one can first collect decision-making criteria and, by creating a hierarchical structure, weigh them so that one can make decisions based on the priority and importance of those criteria [36, 37].

In the following article, after collecting the criteria used in the AHP method, the researcher first collects the current precise location of BTS antennas based on the GIS system, and then, uses the GA as a meta-heuristic method and combines it with the output of the AHP method, examines new candidate locations for the placement of BTS antennas and compares their candidates with the current location locations. The researcher concludes that in the new method, the proposed locations identified in the output of the GA can be more useful and valuable than the current locations because, in general, it improves the basic criteria of the problem, namely, those related to telecommunication issues, those related to the construction and operation of facilities, those related to legal and regulatory issues, and those related to urban issues [35].

The discovery and allocation of optimal locations depend on the goals and parameters of each problem, making their identification a key strength in location allocation studies. This research effectively identifies the features and parameters needed for determining the exact location of BTS, making it a reliable resource for assessment. To compare optimization algorithms, defining a clear and specific objective function is crucial, as it enhances research quality and eliminates inefficient solutions. Objective functions in real-world optimization problems are sometimes complex and difficult, but expressing them and specifying the nature of the functions can determine the research progress and eliminate inappropriate and inefficient solutions. Therefore, in the continuation of research in the field of location allocation, specifying the objective functions and their natures, such as decreasing or increasing, can be fruitful in achieving optimal solutions. In this study, as in other studies, the use of multi-objective optimization algorithms can be suggested to improve the solution based on the characteristics and effective parameters.

Other important points in comparing optimization methods include the number of iterations of the used algorithm, the initial population, or the time to reach optimal solutions. Examining these points and presenting algorithms and methods that reach optimal solutions with fewer parameters in the aforementioned cases are considered strengths. Continuing the current research and comparing it with other optimization algorithms with fewer iterations or initial populations can be a good option for do. Using other optimization algorithms to solve location allocation problems is recommended because of their competitive advantages compared to other methods such as GA [38].

*4) Optimized location allocation of earthquake relief centers using PSO and ACO, complemented by GIS, clustering, and TOPSIS*

Given that most of the world's population lives in areas prone to natural disasters, a lot of research has been recently conducted on disaster management and response conditions [39, 40]. Also, one of the important aspects of urban management is planning for earthquake management, and this starts from the first moment of the earthquake occurrence by starting the relief phase. In the present study, Iran was investigated as a case study and was selected considering that this country is also in an active seismic region with a high number of earthquakes [41]. In this article, in general, after referring to some metaheuristic algorithms such as the crow search algorithm (CSA), the dragonfly algorithm, the whale algorithm [42-44], and... algorithms such as particle swarm optimization (PSO), and ACO, due to the popularity of PSO

[45] and ACO [46] algorithms, these two algorithms were used for optimization.

The performance and features of discrete PSO and ACO algorithms are introduced and examined in this paper, along with the usage of GIS tools to gather urban data, such as in Tehran. Based on factors like land use, slope, acreage, population, distance from fault lines, and roadways, researchers examined active relief locations. Their goal was to use the TOPSIS approach for multi-criteria decision-making to choose 9 locations and allocate areas to them. By maximizing the distance from the negative ideal solution and reducing the distance to the positive ideal solution, TOPSIS finds the optimal solution. Making a decision matrix, normalizing and weighting it, identifying positive and negative ideal solutions, figuring out distances using the Euclidean norm, and ranking according to how close the ideal answer is are all steps in the process. [47]. The researchers in the aforementioned article also divided their identified sites into several clusters using a simple clustering method and selected one site as a representative of each cluster using TOPSIS. The reason for using TOPSIS compared to methods such as AHP or VICOR is the high number of sites for ranking (3065 sites), which seems to be the only TOPSIS method that can be used. In their continuation, they optimized the location of centers by using both ACO and PSO algorithms and compared the outputs with each other. Using 39 sites that were the result of TOPSIS and clustering output, they identified 9 sites for optimal allocation of regions based on these two algorithms. According to the researchers of this study, the PSO algorithm was considered better compared to the ACO algorithm because, for example, it converges faster [48].

Positive features such as comprehensive comparison, comparison with existing methods in GIS, and statistical analyses, make this research a suitable option for review and comparison. Along with these strengths, there are shortcomings such as not providing a specific objective function for comparing methods and limiting the comparison to two optimization algorithms. To improve this research, a specific criterion for comparison can be provided by expressing a mathematical criterion and comparing it with other optimization algorithms to improve the quality of the results obtained and increase its value [49-52].

*5) Geological behavior (GBR) a GIS-based optimization for the fuel location allocation to face the probable tsunami: case of Padang city*

Some areas on Earth, such as the city of Padang in West Sumatra, are exposed to natural disasters such as tsunamis, floods, earthquakes, and the like, which include tsunamis with very high heights of up to 10 meters in that area, which are the result of earthquakes on the seabed due to the movement of the Earth's plates and layers, researchers decided to investigate and present a solution to determine the optimal location of fuel distribution sites, or gas stations, during these natural events so that fuel, which is one of the most vital needs of all individuals and places such as hospitals and refugee camps, can be easily provided to others in times of emergency [53]. Also, a city like Padang faces serious challenges during a tsunami based on its conditions such as distance from the epicenter, flat topography, number of settlements, and economic sites around the coastline [54]. The most important tasks that need to be done against such events are to increase the level of preparedness to succeed in providing immediate assistance at the time of the event and to establish appropriate logistical activities [55]. In the method mentioned in this article, with the help of the ArcGIS application, first, the candidate location for facilities and the probability of road closures based on the worst-case flood are spatially analyzed. Then, using the network analyzer capability, the P-Median method is used to find suitable facilities to meet the stored fuel needs for 72 hours.

The P-Median method is a mathematical optimization algorithm that is used for the problem of determining the optimal location of p service centers such as hospitals, fire stations, etc. In this method, the optimal location of p service centers should be considered such that the total distance between these centers and different locations is minimized. This means that the optimal location should be selected such that the average distance between it and other points is minimized. This algorithm is widely used in problems related to goods distribution, transportation, urban planning, and other such cases [56, 57].

In this study, the researchers first obtained the locations of fuel stations, hospitals, and refugee camps using GIS systems. They then defined a worst-case scenario to identify candidate locations for facilities and potential difficult routes to avoid during the distribution of relief items. They then selected camps and hospitals as target locations because access to fuel was critical to their operations.

They then recorded all the cases in ARC GIS software and presented their solution using the TABU search algorithm (TSA) which is a metaheuristic optimization algorithm [58]. This algorithm is combined with P-Median analysis in the Network Analyst plugin that can be used in ARC GIS software [59], and the researchers used it to present new location positions. Finally, this paper considered the expected bottlenecks in calculating the location allocation solutions. It recommended 9 strategic facilities to serve 18 demand locations with a total fuel volume of 187,660 liters [53]. Finally, the authors suggest their approach can be very effective when integrated with a decision support system (DSS).

According to a quick review of the aforementioned research, the network analysis plugin in ArcGIS software with the TSA calculates facility location allocation points. Error reduction strategies in analytical systems use TSA popularly. Though advantageous, TSA may not be the best algorithm. Applying different optimization methods and comparing them to the suggested approach can enhance it and produce a more reliable answer. For analytical issues, neural networks with error-reduction applications are another possibility. Due to their full connectivity and neuron input weights, these networks may be useful. Like network analysis plugins, neural networks utilize optimization methods to eliminate mistakes, making them ideal for analysis and comparison with this research's method [49-52].

*6) Optimization of garbage collection using genetic algorithm*

In the era of digital transformation, one of the fundamental aspects of making cities smart is optimizing their performance. According to estimates in 2018, public funds will invest $158 billion in smart city initiatives by 2022, nearly twice the $95.8 billion that was anticipated in 2019 [3]. Among the most fundamental issues of investment in smart cities, more complex issues such as urban cleanliness and factors affecting pollution can also be considered. In this regard, the issue of urban waste collection is also discussed and addressed in this article to be more successful in achieving a smart city. In this study, due to the low performance of Brazil in the waste collection and disposal index, as well as problems such as bad odor, visual pollution, and negative issues for public health that need to be addressed, with the help of a favorable optimization, the costs of each region's municipalities for these issues should be minimized [60]. In the present paper, the focus is on discovering the most optimal routes for garbage trucks to minimize pollution and take an important step towards achieving a bluer and cleaner sky by reducing $CO_2$ [61]. This article mentions that previously, using mathematical methods, researchers tried to find optimal routes for garbage truck [62], but they did not evaluate these solutions as appropriate and acknowledged that using newer methods that can take into account several variables and have less computational time can be more beneficial. They further point out that with the increasing expansion of the Internet of Things and Internet and network platforms, it is possible to consider a number of sensors in garbage bins and monitor them, for example, continuously, so that bins that are less than a certain level can be emptied, and in a way, this technique optimizes the movement route of garbage trucks in a more dynamic way. However, due to the smartness of cities, they also considered themselves obliged to receive accurate and recent traffic data in order to use this data in addition to the data received from maps and sensors in their current proposed method.

In this article, the researchers generally used the GA, but in more detail, they suggested that they first use a function to obtain details from the map, such as receiving and using maps from OpenStreetMap (which is a kind of software that provides geographic data to users for free [63]), and taking the output of the information to the Matlab software to perform the main calculations there. This is done by parsing the osm file, in which OpenStreetMap generally provides the user with a series of data that the user can separate from each other if desired, and finally separate, for example, data related to direction or street name or postal code, etc., from each other. Then, they used the Dijkstra algorithm as a default to determine the shortest route, and using map parameters in terms of garbage locations and also combining with traffic data, they used the GA to achieve a more optimal route. In this study, the researchers went into great detail about their method and fully explained the process of achieving the optimal solution, from obtaining map details to applying the GA.

According to the authors' findings, this method, which takes into account the volume of garbage bins filled, can optimize the route traveled by garbage trucks by up to fifteen percent, which in turn leads to lower fuel consumption, which results in less air pollution and greater financial benefits for municipalities. They also suggested further investigation of variables that could contribute to greater financial and environmental benefits for the work.

Optimization research on routing and optimum path selection is complicated and diversified. These problems have been solved by various optimization techniques. This research employs GA with changes to find optimum garbage collection spots and pathways by focusing on the time and iterations needed to attain the optimal solution. GA is used because of its broad use and convenience of use of the provided and basic alternatives for addressing these issues, but it performs poorly in terms of iterations and time to obtain the best solution compared to other optimization methods. Modern algorithms are ideal for these types of studies. In this area, the merging of conventional methods with quantum approaches has been adopted owing to its power, efficiency, and accuracy, proving the superiority of new and combined methods over standard optimization algorithms [62].

Due to the uncertainty in routing problems and discovering the best path, examining and implementing multi-objective algorithms is a more suitable option than single-objective methods. To improve the scientific level in solving such problems, hybrid and multi-objective algorithms can be used as a method to reach a better solution. also in the future, reaching the best location for each garbage bin in each area will be needed.

IV. CONCLUSION AND FUTURE WORKS

Today, in the fields of urban management, smart cities, and natural disaster management, the role of optimization is felt more clearly than before. With the increase in complexity and time to solve optimization problems, sometimes traditional mathematical methods have become inefficient solutions in this field, but the emergence of heuristic and meta-heuristic methods is considered a turning point in solving these problems and facing their challenges. What was mentioned earlier is examples of the application of optimization in various conditions with real-world limitations, which is mostly about determining the location of facilities and various solutions in the field of urban management in the era of digital transformation.

A review of various approaches and research shows the effectiveness of heuristic and meta-heuristic optimization methods in the field of urban management and control, which are considered to be an appropriate and dynamic solution to solve these problems. Addressing the weaknesses and improving research papers in this field in order to achieve higher quality and efficiency is a smart requirement. The shortcomings and weaknesses of the approaches discussed in the present study are described below in a brief.

1. In some cases, lack of expression or transparency in objective functions: Expression of the objective and transparency in its description are key points in problem-solving. In reviewed papers, less attention has been paid to expressing the objective and the requirements of the problem in mathematical language. Using mathematics as a common language in modeling the problem and transforming it into an understandable requirement for all methods and algorithms is essential. Expressing the objective function as clearly as possible seriously helps in solving the problem and understanding its nature.

2. Using basic and not very powerful optimization methods and algorithms: The number of heuristic and meta-heuristic optimization algorithms in today's world is very large. The main difference between these algorithms is in their solving power, the time it takes to reach high quality, the usage rate of computer resources, and stable solutions. Accordingly, some of the optimization algorithms mentioned at the beginning, such as the GA or PSO, have become famous due to the passage of time and the large number of studies conducted with them, but they do not have the necessary power to compete with other newer versions or modern optimization algorithms. Therefore, greedy relying on these methods and not using and comparing them with other methods is considered a weakness in solving these problems.

3. Failure to compare results with other optimization methods and algorithms: The multiplicity of problems with different natures is one of the reasons for comparing the solution used with other optimization methods and algorithms. Different problems can be solved well by some methods, but in others, this rule does not necessarily hold. Hence, comparing existing solutions and other solutions can increase the quality of work and the confidence factor of the obtained solutions.

4. Using single-objective optimization algorithms: In today's world, due to the complexity of problems, simply stating a goal or achieving it is not a desirable option. Optimization algorithms have been developed considering the nature of balancing between goals and achieving solutions that cover multiple goals simultaneously. Hence, using multi-objective optimization algorithms as a multi-dimensional solution is a much more suitable option for solving problems of this style in urban management and the approaches used in it.

For future work and approaches that can be used in future research, the use of new optimization algorithms is suggested by expressing an objective function with a specific criterion and comparing it with other existing methods. Also, in problems where there are different objectives and different coefficients that affect decision-making, multi-objective optimization algorithms are suggested. The use of these methods creates a balance of choices based on the degree of importance and increases the available options for solving problems. The use of benchmarks and comparison tables is also suggested in selecting the appropriate algorithm. On the other hand, in order to increase the speed of conclusion in some cases where real-time results are needed in the field of optimization, it is possible to reach a conclusion using embedded systems, especially FPGA-based systems [64]. This will soon become widespread due to the increasing expansion of Internet of Things-related problems to improve productivity.

Our team will publish the extended version of this article as soon as possible, which can review other articles in detail.